\documentclass[aps,prc,twocolumn,groupedaddress,showkeys,showpacs,floatfix]{revtex4-1}
\usepackage{graphicx}
\usepackage{amsmath,amssymb}
\usepackage{enumerate}
\usepackage{color}
\usepackage[dvipdfm,%
bookmarks=true,%
bookmarksnumbered=true,%
bookmarkstype=toc,%
colorlinks=true,%
linkcolor=blue,%
citecolor=blue,%
pdfauthor={Naofumi Tsunoda},%
pdftitle={Nondeg},%
pdfkeywords={effective interaction,nondegenerate}%
]{hyperref}

\def\bra#1{\langle #1 \rvert}
\def\ket#1{\lvert #1 \rangle}
\def\~#1{\tilde{#1}}
\def\Nu#1#2#3{{}^{#2}_{#3}\mathrm{#1}}
\def\e#1{\epsilon_{#1}}
\newcommand\Heff{H_{\mathrm{eff}}}
\newcommand\HBH{H_{\mathrm{BH}}}
\newcommand\tHeff{\tilde{H}_{\mathrm{eff}}}
\newcommand\tHBH{\tilde{H}_{\mathrm{BH}}}

\newcommand\Veff{V_{\mathrm{eff}}}
\newcommand\Vlowk{V_{\mathrm{low}k}}

\newcommand\fmi{\mathrm{fm}^{-1}}
\newcommand\+{^\dagger}

\newcommand\diff{\mathrm{d}}
\newcommand\w{\omega}
\newcommand\Qbox{$\hat{Q}$-box }
\newcommand\Qboxf{$\hat{Q}$-box}
\newcommand\ntlo{$\chi$N$^3$LO }

\begin{document}


\title{Multi-shell effective interactions}
\author{Naofumi Tsunoda}
\affiliation{Department of physics, the University of Tokyo, 7-3-1 Hongo,
Bunkyo-ku, Tokyo, Japan}
\author{Kazuo Takayanagi}
\affiliation{Department of Physics, Sophia University, 7-1 Kioi-cho,
Chiyoda-ku, Tokyo 102, Japan}
\author{Morten Hjorth-Jensen}
\affiliation{Department of Physics and Center of Mathematics for Applications, University of Oslo, N-0316 Oslo, Norway}
\affiliation{National Superconducting Cyclotron Laboratory and Department of Physics and Astronomy, Michigan
  State University, East Lansing, MI 48824, USA}
\author{Takaharu Otsuka}
\affiliation{Department of physics and Center for Nuclear Study,  
the University of Tokyo, 7-3-1 Hongo, Bunkyo-ku, Tokyo, Japan\\
National Superconducting Cyclotron Laboratory,
Michigan State University, East Lansing, MI, 48824, USA}
\date{\today}
\begin{abstract}
\begin{description}

 \item[Background]
	    Effective interactions, either derived from microscopic
	    theories or based on fitting selected properties of nuclei
	    in specific mass regions, are widely used inputs to shell-model
	    studies of nuclei.  The commonly used unperturbed basis functions are given by the harmonic
            oscillator. 
	    Until recently, most shell-model calculations have
	    been confined to a single oscillator shell like the $sd$-shell or the $pf$-shell. 
	    Recent interest in nuclei away from the stability
	    line, requires  however larger shell-model spaces. Since the derivation of microscopic effective interactions has
            been limited to degenerate models spaces, there are both conceptual and practical limits to present shell-model 
            calculations that utilize such interactions. 
 \item[Purpose]
	    The aim of this work is to present a novel microscopic
	    method to calculate effective nucleon-nucleon interactions
	    for the nuclear shell model. Its main difference from
	    existing theories is that it can be applied not only to
	    degenerate model spaces but also to non-degenerate model spaces.
	    This has important consequences, in particular  for 
	    inter-shell matrix elements of effective 
	    interactions.
 \item[Methods]
	    The formalism is presented in the form of many-body
	    perturbation theory based on 
	    the recently developed Extended Kuo-Krenciglowa
	    method. 
	    Our method
	    enables us to microscopically construct effective
	    interactions not only in one  oscillator  shell but
	    also for several oscillator shells.
 \item[Results]
	    We present numerical results using effective interactions 
	    within (i) a single oscillator shell  ( a so-called degenerate model space)
	    like the $sd$-shell and the $pf$-shell,
	    and (ii) two major shells (non-degenerate model space)
	    like the $sdf_7p_3$-shell and the $pfg_9$-shell.
	    We also present energy levels of several
	    nuclei which have two valence nucleons on top of a given
	    closed-shell core.
 \item[Conclusions]
	    Our results show that the present method works excellently
	    in shell-model spaces that comprise several oscillator shells, as
	    well as in a single oscillator shell. We show in particular that the
	    microscopic inter-shell interactions are much more
	    attractive than has been expected by 
	    degenerate perturbation theory. The consequences for shell-model studies are discussed.
\end{description}
\end{abstract}


\maketitle

\section{Introduction}
The nuclear shell model, a sophisticated theory based on the  configuration
interaction method, has been one of the central theoretical 
tools for understanding a wealth of data from nuclear structure
experiments.  
Due to the rapid growth in the dimensionality of the Hilbert space
with increasing degrees of freedom,
we have to work within a reduced Hilbert space, the so-called model space.
Accordingly, we use an effective interaction which is tailored to the
chosen model space. This effective interaction forms 
an essential input to all shell-model studies.
Equipped with modern sophisticated effective interactions,
the shell model has successfully described many properties of nuclei.

There are two main approaches to determine effective interactions
for the nuclear shell model. 
One is based on fitting two-body matrix elements to 
reproduce observed experimental data.
This approach is widely used in nuclear structure studies, and has been rather successful in
reproducing properties of known nuclei
and in predicting  not yet measured 
properties of nuclei.
The other approach is to {\it derive} the effective interaction 
using many-body theories, 
 starting from bare nucleon-nucleon (NN) interactions.

Although the first approach has been widely used with great
success~\cite{Brown1988191,PhysRevC.65.061301, PhysRevC.70.044307,
Poves1981235, Poves:2001fi},
the main goal of  effective interaction theory is to construct and
understand such sophisticated effective interactions starting from the
underlying nuclear forces and so-called {\em ab initio} or first
principle many-body methods. Most microscopic effective interactions,
except for those used in
no-core shell-model studies~\cite{navratil2009,barrett2013,jurgenson2013}, 
are based on
many-body perturbation theory (see for example
Ref.~\cite{HjorthJensen1995125} for a recent review).
The situation, however, is far from being satisfactory.
In spite of several developments in  many-body perturbation theory,
many properties of nuclei 
are still awaiting a proper microscopic description
and understanding.

A standard approach to derive a microscopic effective interactions for the shell model, is
provided by many-body perturbation theory and the 
so-called folded diagrams~\cite{Kuo_springer} approach. Two widely used approaches  are the 
Kuo-Krenciglowa (KK)~\cite{Krenciglowa1974171} and the
Lee-Suzuki (LS)~\cite{Suzuki80} schemes.
These approaches, however, are feasible only with  degenerate
perturbation theory and are thereby constrained to a model space consisting
of typically one  major oscillator shell.
This poses a strong limitation on the applicability of the
theory. Many unstable nuclei require at least two or more major
oscillator shells for a proper theoretical  description.
For example, the physics of nuclei in the so-called island of inversion
is currently explained 
with empirical effective interactions, see for example Ref.~\cite{PhysRevC.70.044307},
defined for a model space consisting of the $sd$-shell and the $pf$-shell.
It is therefore absolutely necessary to establish
a microscopic theory that allows us to construct an effective
interaction for the model spaces composed of several oscillator shells,
starting from realistic nuclear forces.

Recently, the KK and the LS methods have been extended to the
non-degenerate model spaces~\cite{Takayanagi201161,Takayanagi201191}.
In this work, we present the extended KK (EKK) method
in many-body systems,
which allows us to construct a microscopic effective interaction for several  shells.
We shall see that our theory is a natural extension of
the well-known folded-diagram theory of Kuo and his collaborators
(see for example Refs.~\cite{Krenciglowa1974171,Kuo_springer}).

This article  is structured as follows.
In Sec.~\ref{sec:theory}, 
we briefly explain the concept of 
the effective interaction in a given model space.
In Secs.~\ref{sec:formal} and \ref{sec:MB}, 
we explain our EKK theory for effective interactions. 
We discuss in some detail the difference between the EKK method and
the conventional KK approach which applies  to degenerate model
spaces only.
In Sec.~\ref{sec:test} 
we present test calculations and discussions.
Here we construct effective interactions 
for the nuclear shell model in a single-major
shell ($sd$-shell, $pf$-shell)
and also in two major shells ($sdf_7p_3$-shell, $pfg_9$-shell).
We then calculate energy levels of several nuclei 
that have two valence nucleons on top of
a closed-shell core.
We demonstrate that our method establishes one possible way to reliably
compute microscopic effective interactions for model spaces composed of
several major oscillator shells.
In Sec.~\ref{sec:conclusion} we give a brief conclusion and a summary.

\section{Effective interaction in model space}
\label{sec:theory}

In this section we review briefly the formalism for deriving
an effective interaction using many-body perturbation theory. 

\subsection{model space}\label{sec:veff}

Suppose we describe a quantum system by the following Hamiltonian
\begin{equation}
 H = H_0 + V,
  \label{eq:Hamiltonian}
\end{equation}
where $H_0$ is the unperturbed Hamiltonian and $V$ is 
the perturbation.
In a Hilbert space of dimension $D$, 
 we can write down the many-body Schr\"odinger equation as
\begin{equation}
 H\ket{\Psi_\lambda} = E_\lambda \ket{\Psi_\lambda},\,\,\,\,\,\,
  \lambda = 1, \cdots ,D. \label{eq:schr}
\end{equation}
In shell-model calculations, however, the dimension $D$ of
the Hamiltonian matrix increases exponentially
with the particle number, limiting thereby the applicability of direct diagonalization
procedures to the solution to Eq.~\eqref{eq:schr}.

In this situation, 
we introduce a $P$-space (model space) of a tractable dimension $d\le D$
that is a subspace of the large Hilbert space of
dimension $D$.  
Correspondingly, we define the projection operator $P$ onto the
$P$-space, and $Q=1-P$ onto its complement. We require that
the projection operators $P$ and $Q$ commute with the unperturbed
Hamiltonian $H_0$,
\begin{equation}
 [P,H_0]=[Q,H_0]=0.
\end{equation}

\subsection{Energy-dependent approach}\label{sec:E-dep}

We start our explanation  by 
introducing an energy-dependent effective
Hamiltonian. By use of the projection operators $P$ and $Q$, we can
express Eq.~\eqref{eq:schr} in the following
partitioned form $(\lambda=1,\cdots,D)$:
\begin{equation}
 \begin{pmatrix} PHP & PVQ \\ QVP & QHQ \end{pmatrix}
 \begin{pmatrix} 
  \ket{\phi_\lambda} 
   \\ \ket{\Psi_\lambda}-\ket{\phi_\lambda} 
 \end{pmatrix}
 = E_\lambda
  \begin{pmatrix} \ket{\phi_\lambda} \\ \ket{\Psi_\lambda}-\ket{\phi_\lambda}
  \end{pmatrix},
\label{eq:schr_part}
\end{equation}
where $\ket{\phi_\lambda}=P\ket{\Psi_\lambda}$ is the projection of the
true eigenstate $\ket{\Psi_\lambda}$ onto the $P$-space. Then we can
solve Eq.~\eqref{eq:schr_part}  for $\ket{\phi_\lambda}$ as
\begin{equation}
 \HBH(E_\lambda)\ket{\phi_\lambda}=E_\lambda \ket{\phi_\lambda},\,\,\,\,\,
\lambda = 1, \cdots ,D.
 \label{eq:schr_BH}
\end{equation}
where we have introduced the following Bloch-Horowitz 
effective Hamiltonian $\HBH$ defined 
purely in the $P$-space
\begin{equation}
 \HBH(E)=PHP-PVQ\frac{1}{E-QHQ}QVP.
  \label{eq:def_BH}
\end{equation}
Note that Eq.~\eqref{eq:schr_BH} requires a self-consistent solution, 
because $\HBH(E_\lambda)$ depends on the eigenenergy $E_\lambda$.
This is not a desirable property for the shell-model calculation,
and therefore we adopt the energy-independent approach below.

\subsection{Energy-independent approach}\label{sec:E-indep}

Next we introduce the energy-independent effective Hamiltonian in the
$P$-space. We first choose $d$ eigenstates
$\{\ket{\Psi_i},\,i=1,\cdots,d \}$ among $D$ solutions of
Eq.~\eqref{eq:schr}, with $d\le D$.
Then we require that $\ket{\phi_i}=P\ket{\Psi_i}$, the $P$-space
component of the chosen $d$ eigenstates, be described by the
$d$-dimensional effective Hamiltonian $\Heff$ as
\begin{equation}
 \Heff\ket{\phi_i}=E_i\ket{\phi_i},\,\,\,\,\,i=1,\cdots,d.
\label{eq:schr_eff}
\end{equation}

The energy-independent effective Hamiltonian $\Heff$ can be obtained by
considering the following similarity transformation of the Hamiltonian $H$:
\begin{equation}
 \mathcal{H}=e^{-\w}He^{\w}, \,\,\,\,\,  Q\w P = \w.
\label{eq:sim_trans}
\end{equation}
By construction, the transformed Hamiltonian, $\mathcal{H}$, gives the
same eigenenergies as the original Hamiltonian $H$.
The corresponding eigenstates $\ket{\Psi_i}$, however, are
transformed into $e^{-\w}\ket{\Psi_i}$.
 We require therefore that the second relation 
 in Eq.~\eqref{eq:sim_trans},
  $Q\w P = \w$,
satisfies
$Pe^{-\w}\ket{\Psi_i}=P(1-\w)\ket{\Psi_i}=\ket{\phi_i}$, that is, 
the transformation does not change 
the $P$-space component $\ket{\phi_i}$ of the eigenstates.

Our next step includes the determination of $\w$. 
The most convenient way  to 
determine $\w$ is by using the  following equation
\begin{equation}
 0=Q\mathcal{H}P=QVP-\w PHP+QHQ\w - \w PVQ\w, \label{eq:decoupling}
\end{equation}
which decouples the $P$-space part in the transformed
Schr\"odinger equation. This means that the $P$-space part of the
transformed Hamiltonian, $P\mathcal{H}P$, is nothing but $\Heff$ in
Eq.~\eqref{eq:schr_eff}. Then the effective Hamiltonian and the
effective interaction can be written as 
\begin{equation}
 \Heff=PHP+PVQ\w, \,\,\,\,\,\, \Veff=PVP+PVQ\w.
 \label{eq:def_veff}
\end{equation}
We note here that  $\Heff$ is energy-independent. Furthermore, 
the derivation of  $\Heff$  requires the 
 determination of  $\w$ in order to satisfy
Eq.~\eqref{eq:decoupling}.

\section{Formal theory of effective interaction }\label{sec:formal}

The  decoupling equation \eqref{eq:decoupling}, being nonlinear, can be
solved by iterative methods, to give $\Heff$ and $\Veff$ of
Eq.~\eqref{eq:def_veff}. We first explain the KK
method~\cite{Krenciglowa1974171} for the degenerate model space, and
then turn to the EKK method~\cite{Takayanagi201161,Takayanagi201191} for the
non-degenerate model space. Both methods eliminate the
energy-dependence of $\HBH(E)$ of Eq.~\eqref{eq:def_BH} by introducing
the so-called \Qbox and its energy derivatives, resulting in 
an energy-independent effective interaction $\Heff$.

\subsection{Kuo-Krenciglowa (KK) method}

In the KK method, we assume a degenerate model space, 
$PH_0P=\e 0 P$. Then Eq.~\eqref{eq:decoupling} reads
\begin{equation}
 (\epsilon_0 - QHQ) \w = QVP - \w PV P - \w PVQ \w.
\label{eq:KK1} 
\end{equation}
One way to solve this non-linear equation is to write it in
the following iterative form:
\begin{equation}
 \omega^{(n)}=\frac{1}{\e0-QHQ}\left(QVP-\omega^{(n)} \Veff^{(n-1)}\right),
\end{equation}
where $\omega^{(n)}$ and $\Veff^{(n)}=PVP+PVQ\omega^{(n)}$ stand for
$\omega$ and $\Veff$ in the $n$-th step, respectively. Then we
immediately arrive at the following iterative formula for 
$\Veff^{(n)}$:
\begin{equation}
\Veff^{(n)}=\hat{Q}(\epsilon_0)+
\sum_{k=1}^{\infty}\hat{Q}_k(\epsilon_0)\{\Veff^{(n-1)} \}^k,\label{eq:KK_it}
\end{equation}
where we have defined \Qbox and its derivatives as follows:
\begin{align}
 &\hat{Q}(E)=PVP+PVQ\frac{1}{E-QHQ}QVP, \notag \\
 &\hat{Q}_k(E)=\frac{1}{k!}
  \frac{\diff^k \hat{Q}(E)}{\diff E^k}.
 \label{eq:Qbox}
\end{align}
In the limit of $n \to \infty$, Eq.~\eqref{eq:KK_it} gives 
 $\Veff = \Veff^{(\infty)}$, if the iteration converges.

We stress here that the above KK method can only be applied, by
construction, to a system with a degenerate unperturbed model space
that satisfies $PH_0P=\e 0 P$.
It cannot be applied, for instance, to obtain the effective interaction
for the model space composed of the $sd$-shell  and the $pf$-shell. 

\subsection{Extended Kuo-Krenciglowa (EKK) method}
The extended Kuo-Krenciglowa (EKK) method is designed to construct an
effective Hamiltonian $\Heff$ for non-degenerate model
spaces~\cite{Takayanagi201161,Takayanagi201191}. 
We first rewrite Eq.~\eqref{eq:decoupling} as
\begin{equation}
 (E - QHQ) \w = QVP - \w P\tilde{H}P - \w PVQ \w,\label{eq:EKK_dec}
\end{equation}
where
\begin{equation}
\tilde{H}=H-E
\end{equation}
is a shifted  Hamiltonian obtained by the introduction of the energy parameter $E$.
Equation~\eqref{eq:EKK_dec} plays the same role 
in the EKK method as  Eq.~\eqref{eq:KK1} does in the KK method.
By solving Eq.~\eqref{eq:EKK_dec} iteratively as in the KK method,
we obtain the following iterative scheme to calculate the effective
Hamiltonian $\Heff$,
\begin{equation}
\tHeff^{(n)}=\tHBH(E)+
\sum_{k=1}^{\infty}\hat{Q}_k(E)\{\tHeff^{(n-1)} \}^k, \label{eq:EKK_it}
\end{equation}
where
\begin{equation}
\tHeff=\Heff-E, \,\,\,\,\, \tHBH (E) =\HBH (E) -E,
\end{equation}
and $\tHeff^{(n)}$ stands for $\tHeff$ at the $n$-th step. 
The effective Hamiltonian $\Heff$ is obtained as $\Heff =
\Heff^{(\infty)}$, and satisfies 
\begin{equation}
\tHeff=\tHBH(E)+
\sum_{k=1}^{\infty}\hat{Q}_k(E)\{\tHeff \}^k. \label{eq:EKK_it2}
\end{equation}
The effective interaction, $\Veff$, is then
calculated by Eq.~(\ref{eq:def_veff}) as
$\Veff = \Heff-PH_0 P$.

Let us now compare the EKK and the KK methods.
First, and most importantly, the above EKK method does not require
that the model space be degenerate. It can, therefore, be applied
naturally to a valence space composed of several  shells.
Second, Eq.~\eqref{eq:EKK_it} changes   $\tHeff$, while
Eq.~\eqref{eq:KK_it} changes only $\Veff$ at each step of the iterative
process. 
Third, in order to perform the iterative step of Eq.~\eqref{eq:EKK_it},
we need to calculate $\hat{Q}_k(E)$ at the arbitrarily specified energy
$E$, instead of at $\e0$ for Eq.~\eqref{eq:KK_it}.

Equation~\eqref{eq:EKK_it2} is interpreted as the Taylor series
expansion of $\tHeff$ around $\tHBH(E)$, and changing $E$ corresponds to
shifting the origin of the expansion, and therefore to a re-summation of
the series. This explains why the left hand side of
Eq.~\eqref{eq:EKK_it2} is independent of $E$,
while each term on the right hand side depends on $E$.
This in turn means that we can tune the parameter $E$
in Eq.~\eqref{eq:EKK_it2} to accelerate the convergence 
of the series on the right hand side,  a feature  
which we will exploit in actual calculations.

\section{Many-body  theory of effective interaction}{\label{sec:MB}}
For the purpose of obtaining the effective interaction for the shell
model, we need to apply the formal theory of the effective interaction
in Sec.~\ref{sec:formal} to nuclear many-body systems.
In Sec.~\ref{sec:KK_MB}, we explain the standard Kuo-Krenciglowa (KK)
theory in a many-body system.
Its diagrammatic expression can be established both in
time-dependent~\cite{Kuo_springer} and
time-independent~\cite{Brandow1967} perturbation theory,
and is conveniently summarized  by the $\hat{Q}$-box
expansion in terms of the so-called folded
diagrams~\cite{Kuo_springer}.
In Sec.~\ref{sec:EKK_MB}, we show that $\Heff$ in the EKK method has a
similar expansion 
which can also be expressed by folded
diagrams~\cite{Kuo_springer}.

\subsection{Model space in many-body system}
{\label{sec:model_MB}}

Our quantum many-body system is described by
\begin{eqnarray}
 H &=& H_0 + V \notag\\
 &=& \sum \e {\alpha} a\+_{\alpha} a_{\alpha} 
 + \frac{1}{2}\sum V_{{\alpha}{\beta},{\gamma}{\delta}}
  a\+_{\alpha} a\+_{\beta} a_{\delta} a_{\gamma},
  \label{eq:MB_Hamiltonian}
\end{eqnarray}
where $H_0$ is the unperturbed Hamiltonian and $V$ is the two-body
interaction. We limit ourselves, for the sake of simplicity, to two-body
interactions only, although the theory can be extended to include 
three-body or more complicated nuclear forces.

In specifying single-particle states, we use indices $a,b,c,d$ for
valence single-particle states (active single-particle states), and $p$ and $h$ for passive particle
and hole single-particle states, respectively. 
In a generic case, we use Greek indices.

In a many-body system, the $P$-space is defined using the valence
single-particle states that make up the $P$-space. Let us take as an example the nucleus
$\Nu{O}{18}{}$, where we treat $\Nu{O}{16}{}$ as a closed-shell core.
In this case we can define the $P$-space
by specifying the valence states to be determined  by the single-particle states of the $sd$-shell.
The $P$-space is then composed by the $\Nu{O}{16}{}$ closed-shell core plus two
neutrons in the $sd$-shell.

In the following, we derive $\Heff$ (and $\Veff$) in
Eq.~(\ref{eq:def_veff}) with the above Hamiltonian
Eq.~\eqref{eq:MB_Hamiltonian}, which gives
\begin{equation}
 \Heff\ket{\phi_i}=E_i\ket{\phi_i},\,\,\,\,\,i=1,\cdots,d,
\label{eq:schr_effKK}
\end{equation}
for the many-body system.

\subsection{Kuo-Krenciglowa (KK) method}{\label{sec:KK_MB}}
Here we briefly explain the KK method in a degenerate $P$-space.
Many body perturbation theory (MBPT) shows that the effective
interaction, $\Veff$, is conveniently written in terms of the so-called
folded diagrams as~\cite{Kuo_springer}
\begin{equation}
 \Veff=\hat{Q}(\e0)-\hat{Q'}(\e0)\int\hat{Q}(\e0)+
  \hat{Q'}(\e0)\int\hat{Q}(\e0)\int\hat{Q}(\e0)\cdots,
  \label{eq:Veff_Qbox_expansion}
\end{equation}
where the integral sign  represents  the folding procedure, and $\hat{Q'}$
represents \Qbox contributions which have at least two nucleon-nucleon
interaction vertices.
Note that, in order to have a degenerate $P$-space energy, $\e0$,
the single-particle energies in Eq.~\eqref{eq:MB_Hamiltonian} for 
valence single-particle states, $\e {a},~\e {b},\dots$ are completely degenerate.
Equation~\eqref{eq:Veff_Qbox_expansion} is the basis of the perturbative
expansion of $\Veff$ in the folded diagram theory (see for example
Ref.~\cite{Kuo_springer} for more details).

\begin{figure}[htbp]
 \begin{center}
  \includegraphics[width=5cm,clip,angle=0]{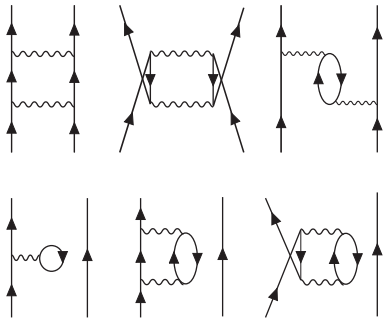}
  \caption{Valence-linked \Qbox diagrams up to second
  order in $V$.}\label{fig:diagram}
 \end{center}
\end{figure}

There are two points to be noted here. 
First, because we cannot evaluate the \Qbox defined in
Eq.~\eqref{eq:Qbox} exactly (which implies including all terms to
infinite order), we use the following perturbative expansion
\begin{eqnarray}
 \hat{Q}(E)&=&PVP+PV\frac{Q}{E-H_0}VP \nonumber\\
  &&+ PV\frac{Q}{E-H_0}V \frac{Q}{E-H_0}VP + \cdots,
 \label{eq:Qbox_expansion}
\end{eqnarray}
which we can currently evaluate up to the third order in the
nucleon-nucleon interaction. 
Second, the valence-linked diagram theorem states that we need to
retain only the valence-linked part (See Fig.~\ref{fig:diagram}), i.e.,
unlinked parts can be proved to cancel
among themselves ~\cite{Kuo_springer,Brandow1967}.
At the same time, the eigenvalue $E_i$ in Eq.~\eqref{eq:schr_effKK}
changes its meaning; 
it is no longer the total energy of the system,
but is now the total energy measured from the true ground state energy
of the core.

In actual calculations, however, 
we do not calculate $\Veff$ order by order using 
Eq.~\eqref{eq:Veff_Qbox_expansion}.
Since  the contribution of folded diagrams 
can be calculated by energy derivatives when
the model space is degenerate~\cite{Kuo_springer},
we can translate Eq.~\eqref{eq:Veff_Qbox_expansion}
into the following equation
\begin{equation}
\Veff=\hat{Q}(\epsilon_0)+
\sum_{k=1}^{\infty}\hat{Q}_k(\epsilon_0)\{\Veff \}^k,
\label{eq:Veff_Qbox_expansion2}
\end{equation}
The above expression clearly shows that 
the iterative solution of Eq.~\eqref{eq:KK_it} converges
$\Veff$ in the limit of $n \to \infty$.

We can summarize the KK method as follows;  
we calculate the valence-linked \Qbox diagrams (usually up to second or
third order) and the corresponding energy derivatives at the degenerate
$P$-space energy $\e0$, and carry out the iteration of Eq.~\eqref{eq:KK_it}
starting from $\Veff^{(0)}=V$.
This procedure ultimately gives $\Veff= \Veff^{(\infty)}$.

At the end, we stress again that the above KK method can yield $\Veff$
only for a  degenerate model space. Suppose we are working with the
harmonic oscillator shell model of $\Nu{O}{18}{}$, treating
$\Nu{O}{16}{}$ as the core. If we take the $P$-space composed
only of the degenerate $sd$-shell, the above KK method works well 
 as shown by many applications
  (see for example Ref.~\cite{HjorthJensen1995125}).
If, on the other hand, we take an enlarged $P$-space defined by the
non-degenerate $sdf_7p_3$-shell, the KK method breaks down.
A naive calculation of $\Veff$ by Eq.~\eqref{eq:Veff_Qbox_expansion}
easily leads to divergences of the \Qboxf, as we shall see later.

\subsection{Extended Kuo-Krenciglowa method}\label{sec:EKK_MB}
Here we derive the effective Hamiltonian $\Heff$
 of the Extended Kuo-Krenciglowa (EKK) method, 
 with an emphasis on its similarity with the KK
method discussed above.

\subsubsection{Derivation of the Extended Kuo-Krenciglowa
method}\label{sec:EKK_MB_deriv}

We consider first the general situation where the energies
of the valence single-particle states in $P H_0 P$ are not necessarily degenerate.
In this case, we have to apply the EKK formula Eq.~\eqref{eq:EKK_it} to
our many-body systems.
 
We can easily confirm that, in order to derive Eq.~\eqref{eq:EKK_it},
we  need to change the decomposition  
Eq.~\eqref{eq:MB_Hamiltonian} of the Hamiltonian in the KK method. 
Suppose we decompose the total Hamiltonian into 
the following unperturbed Hamiltonian $H'_0$ and
the perturbation $V'$
\begin{eqnarray}
 H'_0&=&PEP+Q H_0 Q \notag\\
 V'&=&V- 
 P(E-H_0) P,
  \label{eq:EKK_H}
\end{eqnarray}
or in the matrix form,
\begin{eqnarray}
 H&=&H'_0+V' \notag \\
 &=&
  \begin{pmatrix}
   E &0 \\ 0 & QH_0Q
  \end{pmatrix} + \begin{pmatrix} 
   P\tilde{H}P & PVQ \\ QVP &QVQ
  \end{pmatrix},
  \label{eq:EKK_H_mat}
\end{eqnarray}
where $\tilde{H}\equiv H-E$, and $H_0=\sum \e {\alpha} a\+_{\alpha}
a_{\alpha}$.
With the above unperturbed Hamiltonian $H'_0$ in Eq.~\eqref{eq:EKK_H},
we can treat the $P$-space as being  degenerate at the energy $E$,
and therefore we can follow the derivation of
Eq.~\eqref{eq:Veff_Qbox_expansion} in the KK method, to achieve 
\begin{equation}
 \tHeff=
  \tHBH (E)-\hat{Q'} (E)\int\tHBH (E)+
  \hat{Q'} (E)\int\tHBH (E)\int\tHBH (E)\cdots, 
  \label{eq:Heff_Qbox_expansion}
\end{equation}
\noindent
which is then converted into
\begin{equation}
 \tHeff=\tHBH (E)
  + \frac{\diff \hat{Q}(E)}{\diff E} \tHeff+
  \frac{1}{2!}\frac{\diff^2 \hat{Q}(E)}{\diff E ^2} \{\tHeff\}^2 + \cdots.
\label{eq:Heff_Qbox_expansion2}
\end{equation}
Note that Eqs.~\eqref{eq:Heff_Qbox_expansion} and
\eqref{eq:Heff_Qbox_expansion2} are  the EKK counterparts of
Eqs.~\eqref{eq:Veff_Qbox_expansion} and \eqref{eq:Veff_Qbox_expansion2}
of the KK method. We can solve Eq.~\eqref{eq:Heff_Qbox_expansion2}
iteratively as done for Eq.~\eqref{eq:Veff_Qbox_expansion2}. 

Note that the above derivation of the EKK method is the same as that of 
the KK method apart from the decomposition of the Hamiltonian.
Therefore, we need to retain only
the valence-linked \Qbox diagrams in Eqs.~\eqref{eq:Heff_Qbox_expansion}
and \eqref{eq:Heff_Qbox_expansion2}, as we do for the KK method.
To summarize, all that we have to know is, as in the KK method, the \Qbox
and its energy derivatives, except that now it is defined at the parameter $E$.

\subsubsection{\texorpdfstring{Perturbative expansion of the
$\hat{Q}$-box}{Perturbative expansion of the Q-box}}
\label{sec:EKK_MB_Qbox}

We discuss here how one can accommodate the
perturbative expansion of the $\hat{Q}$-box
in the EKK formalism. For the sake of simplicity,
we will  focus on a simple system composed of  two particles
on top of a closed-shell core 
in what follows, although the discussion is not restricted to this
specific case.
The projection operators $P$ and $Q$ are  then given by 
\begin{equation}
 P=\sum_{a,b}a\+_a a\+_b\ket{c}\bra{c}a_b a_a, \,\,\,\,\,\, Q=1-P.
\end{equation}
where $\ket{c}$ stands for the closed-shell core.

To become familiar with our new unperturbed Hamiltonian, $H'_0$,
we consider some selected examples;
 we show first the results from the 
 operation of the new unperturbed Hamiltonian, $H_0'$, 
 on some selected many-particle states:
\begin{eqnarray}
&& H_0'~a\+_a a\+_b\ket{c}= E~ a\+_a a\+_b\ket{c}, \nonumber \\
&& H_0'~a\+_a a\+_p\ket{c}= (\e a + \e p)~ a\+_a a\+_p\ket{c},  \\
&& H_0'~a\+_a a\+_b a\+_p a_h \ket{c}
  = (\e a + \e b + \e p - \e h)~ a\+_a a\+_b a\+_p a_h\ket{c}. 
 \nonumber
\end{eqnarray}
The first line is an example of a $P$-space state with two
single-particle states on top of the closed-shell core, $|c\rangle$,
while the second and third lines result in $Q$-space examples.
The unperturbed energy of the $P$-space state $a\+_a a\+_b\ket{c}$ is $E$,
while that of a $Q$-space state $a\+_a a\+_p\ket{c}$ results in  
$\e a + \e p$,  a sum of  unperturbed single-particle
energies. It is important to get convinced that $\e a$ appears only
in the $Q$-space energy, while $a\+_a$ appears in all of the above three
states.

In the perturbative expansion of $\hat{Q}(E)$ in
Eq.~\eqref{eq:Qbox_expansion}, all the intermediate states are in the
$Q$-space, and their unperturbed energies are given as in
the second and third lines of the above example. The general structure
of $\hat{Q}(E)$ can then be  given schematically by
\begin{eqnarray}
 \hat{Q}(E)=\prod \frac{V}{E- (\sum\e{a}+\sum\e{p}-\sum\e{h})_{int}} ,
  {\label{eq:EKK_diag}}
\end{eqnarray}
where the subscript $int$ indicates intermediate states between two
interaction vertices. Note that the parameter $E$ appears in all the
denominators in the EKK method.

\begin{figure}[htbp]
 \includegraphics[width=3cm,clip,angle=0]{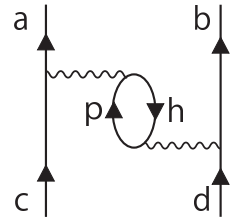}
 \caption{Core-polarization diagram as a second order in $V$
 contribution to the \Qboxf.} \label{fig:cp}
\end{figure}

Let us consider the \Qbox diagram shown in
Fig.~\ref{fig:cp} as an example.
 The diagram is a contribution  from the second-order term in
Eq.~\eqref{eq:Qbox_expansion}, and gives the following
contribution to $\hat{Q}(E)$
\begin{equation}
\mathrm{Fig.2(EKK)} \rightarrow
\frac{V_{ah,cp}V_{pb,hd}}{E-\e{c}-\e{b}-\e{p}+\e{h}}.
  \label{eq:diag_EKK}
\end{equation}
If we on the other hand employ the KK method in order  to calculate
the contribution to $\hat{Q}(\e 0)$ from Fig.~\ref{fig:cp}, we would get
\begin{eqnarray}
\mathrm{Fig.2(KK)} 
& \rightarrow & \frac{V_{ah,cp}V_{pb,hd}}
    {(\e{c}+\e{d})-\e{c}-\e{p}+\e{h}-\e{b}} \nonumber\\
&= & \frac{V_{ah,cp}V_{pb,hd}}
    {-\e{p}+\e{h}}
  \label{eq:diag_KK}
\end{eqnarray}
where, in going to the second line, we have used the fact that the
$P$-space is degenerate, and therefore $\e{a}=\e{b}=\e{c}=\e{d}$ and
$\e{c}+\e{d}=\e0$. 

Two points should be noted from  the above example;
 first, in a degenerate model space,
 the EKK result Eq.~\eqref{eq:diag_EKK} with $E=\e0$
 coincides with the KK result Eq.~\eqref{eq:diag_KK}. 
 This is a direct consequence of the fact that the EKK formula
contains the KK formula as a special case. 
Second, we can see the problem of divergence
of the KK formula applied naively to a non-degenerate model space. Consider  
the case of  $\Nu{O}{18}{}$ as an example, and let the $P$-space consist
of two major shells ($1s0d$ and $1p0f$-shells). Then, the denominator
of the first line in Eq.~\eqref{eq:diag_KK} vanishes for $b,c,p
\in 1s0d$-shell, $a,d \in 1p0f$-shell, and $h \in 0p$-shell,
 leading thereby to a divergence.
 The above mechanism is just one of many examples which show that we really need to use the EKK formalism
 in order to derive effective interactions for model spaces consisting of several shells.

\subsubsection{\texorpdfstring{The energy parameter $E$}
   {The energy parameter E}}
   \label{sec:EKK_MB_E}

Here we make several points  in connection with the new parameter $E$
in the EKK theory.

First, by virtue of the arbitrariness of the parameter $E$,
 we can get around the problems of the poles of the $\hat{Q}$-box.
As explained with the example in Fig.~\ref{fig:cp},
the $\hat{Q}$-box contribution, \eqref{eq:diag_KK}, 
is divergent in the KK formula
when we work in a non-degenerate model space.
On the other hand, the EKK counterpart, \eqref{eq:diag_EKK},
becomes divergent only for specific values of $E$.
This means that we can always select a parameter $E$ that 
makes the $\hat{Q}$-box to behave smoothly as a function of $E$.

Second, the parameter $E$ provides us with a reasonable test of the
calculation. 
Equation~\eqref{eq:EKK_it2} shows that the resulting effective
Hamiltonian $\Heff$, does not depend on the value of $E$,
provided that we have the exact \Qbox in Eq.~\eqref{eq:EKK_it2}.
Any approximation may spoil the $E$-independence of $\Heff$, and
thereby  of other physical quantities described by $\Heff$.
This may in turn imply that a \Qbox that gives $E$-independent physical
quantities is a good approximation to the exact $\hat{Q}$-box.
In the most sophisticated calculations up to date,
we can evaluate the \Qbox diagrams only up to third order in the
perturbation $V$ at most,
and we can include only a limited number of excitations in the
intermediate states (typically including up to  $10-20$ oscillator
quanta excitations), because of practical computational limitations.
In Sec.~\ref{sec:test}, we shall present numerical tests of the above
$E$-independence.

Here we explain how to fix the value of $E$ in the actual
calculations. If there are no intruder states in the target states 
$\{\ket{\phi_i},\,i=1,\cdots,d\}$ that are to be described by $\Heff$,
it can be shown that there is a finite range of $E$ values that make the
series in Eq.~\eqref{eq:Heff_Qbox_expansion2}
convergent~\cite{Takayanagi201191}.
The convergence is usually at its fastest when $E$ is fixed around the mean
value of the target energies, $\{ E_i,\,i=1,\cdots,d \}$. Let us come
back to our specific case of two particles on top of a core, 
and employ a single-particle basis of the harmonic
oscillator.
Let $\e a^{\mathrm{min}}$ be the minimum energy of active (valence)
particle states. 
We then expect that our target states are distributed around $E \sim 2\e
a^{\mathrm{min}}$, which serve as the first guess for $E$.
It is also clear that the lowest energy of the $Q$-space states is 
$2\e a^{\mathrm{min}}+1\hbar \w$, which gives the lowest position of the
poles of $\hat{Q}(E)$.

In actual calculations, the allowed range of $E$, which is limited both
from above and from  below, can be estimated as follows.
Let us increase $E$ from the above ``{\it optimal} ''
value $2\e a^{\mathrm{min}}$.
As $E$ approaches the lowest pole $2\e a^{\mathrm{min}}+1\hbar \w$ of
$\hat{Q}(E)$, $\hat{Q}(E)$ and its derivatives in
Eq.~\eqref{eq:Heff_Qbox_expansion2} would diverge.
On the other hand, if we choose too low a value of $E \ll 2\e
a^{\mathrm{min}}$, the resultant energy denominators in the $\hat{Q}(E)$
would be dominated by $E$, but not by the unperturbed energies of the
intermediate states.
In this situation, we have to expect that our approximation,
truncation of the intermediate states at some unperturbed energies,
cannot be justified.

In the next section, we present numerical examples where $E$ is varied in
 some range around $E \sim 2\e a^{\mathrm{min}} $.

\section{Numerical results}\label{sec:test}
In order to apply the EKK formalism to nuclear
systems, we consider simple nuclei composed of two nucleons on top of a
given closed-shell core, $\Nu{O}{18}{}, \Nu{F}{18}{}$ and
$\Nu{Ca}{42}{}, \Nu{Sc}{42}{}$, employing a single-particle basis
defined by the harmonic oscillator unperturbed Hamiltonian.
As the $P$-space for $\Nu{O}{18}{}$ and $\Nu{F}{18}{}$, we employ the
$sd$-shell (degenerate case), and the $sdf_7p_3$-shell 
(non-degenerate case) composed of the $sd$ shell and the $0f_{7/2}$ and
the $1p_{3/2}$ single-particle states. 
The $P$-space for $\Nu{Ca}{42}{}, \Nu{Sc}{42}{}$ is the $pf$-shell
(degenerate case), and the $pfg_9$-shell (non-degenerate case) composed
of all  the  $pf$-shell single-particle states  and the
$0g_{9/2}$ single-particle state.
In the degenerate $P$-space, both the KK and the EKK methods can be
used, while in the non-degenerate $P$-space only the EKK method is
applicable.

Our input interaction $V$ in the Hamiltonian
Eq.~\eqref{eq:MB_Hamiltonian} is the low-momentum
interaction $\Vlowk$ with a sharp cutoff $\Lambda = 2.5~\fmi$, 
derived from the chiral \ntlo interaction of Entem and
Machleidt~\cite{entem2003,machleidt2011}.
Our total $(P+Q)$ Hilbert space is composed of the harmonic oscillator
basis states in the lowest seven major shells. The $Q$-space degrees of
freedom come into play either by the KK method or the
EKK method through the \Qbox that
is calculated 
to third order in the interaction $V$.
 The final effective interactions are thus obtained in 
 the $P$-space of one or two major shells. 
 
We ought to mention two points. First, the amount of oscillator quanta
excitations in each term in perturbation theory, may not be fully
adequate if one is interested in final shell-model energies that are
converged with respect to the number of intermediate excitations in
the \Qbox diagrams.  Second, neglected many-body correlations, like
those arising from three-body forces are not taken into account. Such
effects, together with other many-body correlations not included here
are expected to play important roles, see for example the recent
studies of neutron-rich oxygen and calcium
isotopes~\cite{PhysRevLett.105.032501,Holt:2010yb,Otsuka:2013ig,
  PhysRevLett.108.242501,PhysRevLett.109.032502}.  However, the aim
here is to study the effective interactions that arise from  the KK and the EKK
methods in one and two major shells.  Detailed calculations for nuclei
along various isotopic chains will be presented elsewhere.

In the actual calculation of the \Qbox using the  harmonic oscillator
basis functions, we drop the Hartree-Fock diagram
contributions, assuming that it is well simulated by the
harmonic oscillator potential, as in many of the former
works~\cite{HjorthJensen1995125}. 

In order to show how the EKK method works, we present separate studies
of (i) the two-body matrix elements (TBMEs) of the effective interaction
$\Veff$, and (ii) several energy levels obtained by shell
model calculations. 
In particular, we focus on the $E$-independence
of the numerical results; as discussed in
Sec.~\ref{sec:theory}, the effective Hamiltonian $\Heff$ 
obtained with the exact \Qbox is independent
on the energy parameter $E$, and so are physical quantities 
calculated with  $\Heff$.
In other words, the explicit $E$-dependence of the first term
$\HBH (E)$ (or equivalently $\hat{Q}(E)$) 
in Eq.~\eqref{eq:Heff_Qbox_expansion2}
is canceled by other terms that
 represent  the folded diagram contributions, making thereby
$\Heff$ (and therefore $\Veff$)
energy independent.
In actual calculations, however, we can calculate the \Qbox only up to
third order in $V$, and we have to examine the $E$-dependence of the
right hand side of  Eq.~\eqref{eq:EKK_it2}.
In what follows, 
we shall see clearly that the \Qbox up to third order is sufficient to
achieve an almost $E$-independent  effective interaction $\Veff$ (or $\Heff$).

\subsection{\texorpdfstring{EKK method in $sd$-shell and $sdf_7p_3$-shell}
{EKK method in sd-shell and sdf7p3-shell}}
\label{sec:sd_sdf7p3}

In this section,
we consider  $\Nu{O}{18}{}$ and $\Nu{F}{18}{}$ as two-nucleon
systems on top of the $\Nu{O}{16}{}$ closed-shell core.
We calculate 
the effective interactions $\Veff$ in the degenerate $sd$-shell, 
and in the non-degenerate $sdf_7p_3$-shell model spaces.
The input of the EKK method is the
 \Qbox calculated up to third order in $V$
 with the harmonic oscillator
basis of $\hbar\w=14~\mathrm{MeV}$.
We set the origin of the energy as $\e{sd} = 0$, 
which suggests that the optimal choice of $E$ is $E \sim 0$.

\subsubsection{\texorpdfstring{degenerate $sd$-shell model space}
{degenerate sd-shell model space}}
\label{sec:sd}
\begin{figure*}[htbp]
  \includegraphics[width=12cm,angle=0,clip]{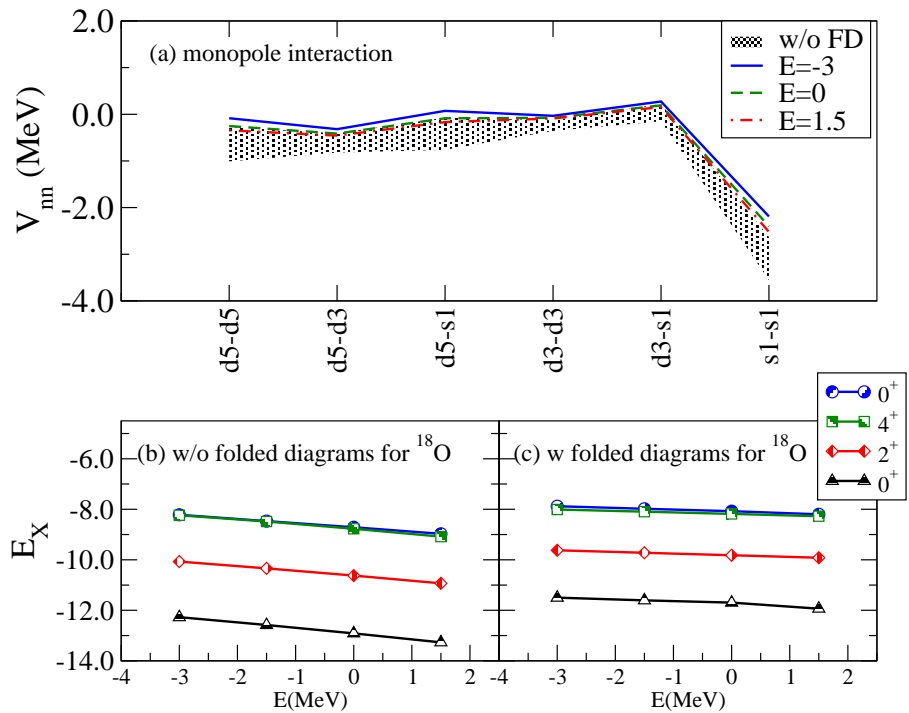}
 \caption
 {(Color online) $E$-dependence of the EKK results of neutron-neutron channel in the
 $sd$-shell (degenerate model space).
 The figure shows the monopole part,
 of $\Veff$ (denoted by $\mathrm{V_{nn}}$), see Eq.~(\ref{eq:monopole}), 
 and the level energies of $\Nu{O}{18}{}$ with respect to $\Nu{O}{16}{}$.
 In panel (a), the monopole panel, dotted lines (which make the shaded area)
 show the results without the folded diagram contributions
 for $-3\leq E \leq 1.5~\mathrm{MeV}$.
 The full line, dashed line and dot-dash line show the $\mathrm{V_{nn}}$
 for $E=-3,~0,~1.5~\mathrm{MeV}$, respectively.
 In the lower panels (b) and (c), energy levels are calculated for 
 $E=-3,~-1.5,~0,~1.5~\mathrm{MeV}$ (b) without and (c) with the folded
 diagram contribution.
 Triangles, diamonds, squares and circles show the energy levels of the ground,
 the first excited, the second and third excited states, respectively.}
 \label{fig:sd_result_nn}
\end{figure*}
\begin{figure*}[htbp]
  \includegraphics[width=12cm,angle=0,clip]{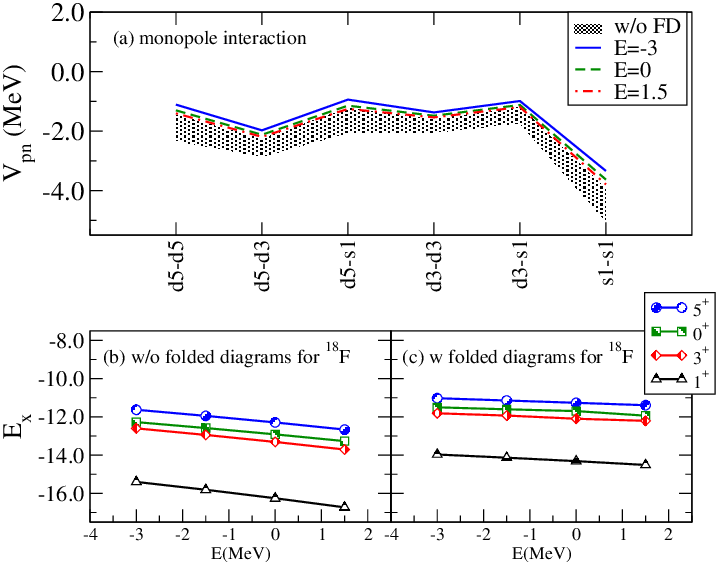}
 \caption
 {(Color online) $E$-dependence of the EKK results of proton-neutron channel
 in the $sd$-shell (degenerate model space).
 (a) monopole part of the interaction, (b) level energies of
 $\Nu{F}{18}{}$ without folded diagrams and (c) with folded diagrams.
 Else, we use the same notation as in Fig.~\ref{fig:sd_result_nn}.}
\label{fig:sd_result_pn}
\end{figure*}
In Fig.~\ref{fig:sd_result_nn}, we show
our numerical results for the two-body matrix elements and level
energies calculated in the degenerate 
$sd$-shell model space for the neutron-neutron channel ($\Nu{O}{18}{}$) 
and in Fig.~\ref{fig:sd_result_pn} the proton-neutron
channel ($\Nu{F}{18}{}$).

To see the $E$-independence of the numerical results for $\Veff$, 
calculations are performed for several values of $E$.
As explained before, 
the optimal value of $E$ may be estimated as $E\sim 2\e {sd}=0$.
Note also that $E=0$ is far from the 
lowest pole of $\hat{Q}(E)$,
$E=E_\mathrm{pole}^{min}=1~\hbar\w=14~\mathrm{MeV}$,
and the calculation is free from the divergence problem of 
the \Qboxf.
We have thus varied $E$ in the range of $-3\leq E \leq 1.5~\mathrm{MeV}$ in
Fig.~\ref{fig:sd_result_nn} and \ref{fig:sd_result_pn}.
Obviously, the EKK method with $E=0$ coincides exactly with the KK method
because our $P$-space is degenerate now (compare Eqs.~\eqref{eq:diag_EKK}
and \eqref{eq:diag_KK}).

In order to study the effect on the various matrix elements, 
we analyze the monopole term in the neutron-neutron channel  in
Fig.~\ref{fig:sd_result_nn} (a). The monopole part of $\Veff$,
is defined as 
\begin{equation}
 \Veff{}_{j,j'}^T=\frac{\sum_{J}(2J+1) \bra{jj'}
  \Veff \ket{jj'}_{JT}}{\sum_{J} (2J+1)}.\label{eq:monopole}
\end{equation}
Let us look at the dotted lines (which make a shaded band in the figure)
that are calculated by dropping the folded diagram contributions,
i.e., by replacing $\Veff$ simply by $\hat{Q}(E)$.
We can see clearly that $\hat{Q}(E)$ depends strongly on $E$.
Next, let us turn to the EKK results that include all the folded diagram
contributions in the right hand side of
Eq.~\eqref{eq:Heff_Qbox_expansion2}.
They are shown by solid lines for $ E=-3,~0,~1.5 ~\mathrm{MeV}$,
whose difference can hardly be seen.
The above observation suggests that the folded diagrams cancel the
$E$-dependence of $\hat{Q}(E)$ and yield an almost
$E$-independent $\tHeff$ (and $\Veff$) in
Eq.~\eqref{eq:Heff_Qbox_expansion2}.

In the lower panels (b) and (c) of Fig.~\ref{fig:sd_result_nn},
we  show several energy levels of $\Nu{O}{18}{}$ with
respect to $\Nu{O}{16}{}$ obtained by shell-model calculations with our
effective interaction $\Veff$.
Here the single particle energies in the shell-model
diagonalization are taken from the USD
interaction~\cite{Brown:1988ff,Brown1988191}; the single-particle energies of
the states (in the isospin formalism) $\e {d_{5/2}},~\e {s_{1/2}},$ and $\e {d_{3/2}}$ are
$-3.9478$ MeV, $-3.1635$ MeV and $1.6466$ MeV, respectively.

Panels (b) and (c) show the results without and with folded
diagram contributions, respectively.
We note that in panel (b) the energy levels 
 are decreasing functions of $E$,
which is explained  by the $E$-dependence of $\hat{Q}(E)$.
On the other hand, in panel (c), we see that the energy levels are almost
independent of the parameter $E$, as they should.

The above observation also implies that the evaluation of the \Qbox
up to third order in $V$ is sufficient to establish a seemingly $E$-independence
of the right hand side of Eq.~\eqref{eq:Heff_Qbox_expansion2},
and therefore of $\Veff$ in the left hand side.

Figure~\ref{fig:sd_result_pn} shows the results for $\Veff$ in
the proton-neutron channel and level energies for $\Nu{F}{18}{}$ with
the same setting as for $\Nu{O}{18}{}$.
We can repeat exactly the same arguments as we did for $\Nu{O}{18}{}$ and realize
that the folded diagram contribution eliminates the $E$-dependence, to a large extent, of
$\hat{Q}(E)$ to give, almost, an $E$-independent $\Veff$ and energy levels.
 
\subsubsection{\texorpdfstring{Non-degenerate $sdf_7p_3$-shell}
{Non-degenerate sdf7p3-shell}}\label{sec:sdf7p3}
Here we examine $\Nu{O}{18}{}$ and $\Nu{F}{18}{}$ in the
non-degenerate $P$-space, labelled the $sdf_7p_3$-shell here, composed
of the $sd$-shell and the $0f_{7/2}$ and $1p_{3/2}$ single-particle
states.  In this non-degenerate model space, the standard KK method
cannot be applied, since it leads to divergences, as discussed
above. The EKK method offers however a viable approach to this system.
To date, therefore, there have been only empirical interactions in
this model space, see for example the effective interaction employed in Ref.~\cite{PhysRevC.60.054315}.

\begin{figure*}[htbp]
 \includegraphics[width=12cm,angle=0,clip]{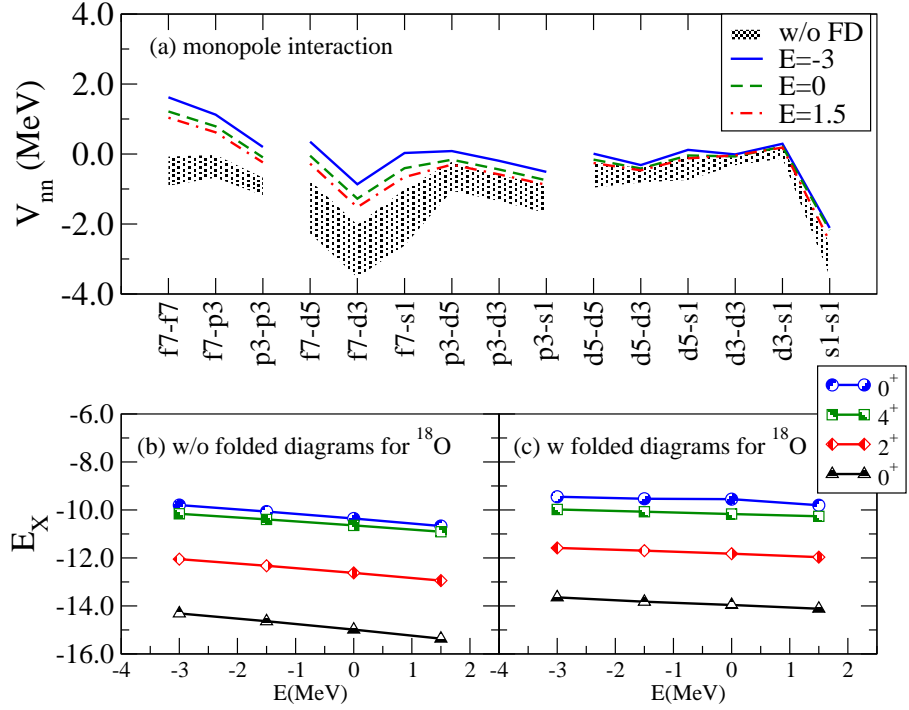}
 \caption{(Color online) $E$-dependence of the EKK results of neutron-neutron channel
 in the $sdf_7p_3$-shell (non-degenerate model space).
 (a) monopole part of the interaction, (b) level energies of
 $\Nu{O}{18}{}$ without folded diagrams and (c) with folded diagrams.
 Else, we use the same  notation as in  Fig.~\ref{fig:sd_result_nn}.}
 \label{fig:sdf7p3_result_nn}
\end{figure*}
\begin{figure*}
 \includegraphics[width=12cm,angle=0,clip]{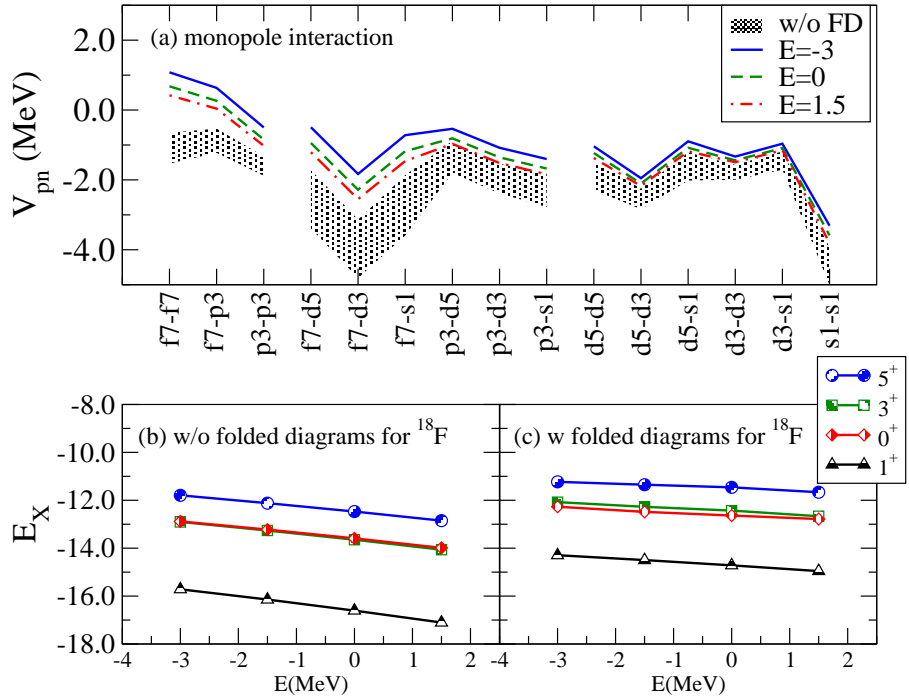}
 \caption{(Color online) $E$-dependence of the EKK results of proton-neutron channel
 in the $sdf_7p_3$-shell (non-degenerate model space).
 (a) monopole part of the interaction, (b) level energies of
 $\Nu{F}{18}{}$ without folded diagrams and (c) with folded diagrams.
 Else, we use the same  notation as in  Fig.~\ref{fig:sd_result_pn}.}
\label{fig:sdf7p3_result_pn}
\end{figure*}

Figures~\ref{fig:sdf7p3_result_nn} and~\ref{fig:sdf7p3_result_pn} 
show the numerical results of the EKK
method in the $sdf_7p_3$ model space, in the same way as
Figs.~\ref{fig:sd_result_nn} and \ref{fig:sd_result_pn} 
for the degenerate $sd$-shell.
In what follows,
we will introduce the wording inter-shell interaction for  the interaction between particles 
in different major shells, for example we can have one particle 
in the $sd$-shell and the other one in the $pf$-shell. Similarly, we will use the naming 
intra-shell interaction for 
interactions between particles within a single major shell.

Let us first study the the monopole part of the TBMEs of $\Veff$
(Fig.~\ref{fig:sdf7p3_result_nn} (a) for neutron-neutron channel and
Fig.~\ref{fig:sdf7p3_result_pn} (a) for proton-neutron channel).
Here we have inter-shell interactions
in addition to the intra-shell interactions within the $sd$-shell and
within the $pf$-shell. We see that the inter-shell interactions depend more
strongly on $E$ without the folded diagrams than the intra-shell
interaction within the $sd$-shell.
This feature is due to the fact that  $\hat{Q}(E)$ 
for the inter-shell interaction
has intermediate states with small energy denominators.
This figure clearly shows that the folded diagram contributions cancel
the strong $E$-dependence of the $\hat{Q}(E)$, and the resulting $\Veff$
becomes almost independent of $E$, showing that the EKK method 
is stable and useful also in the non-degenerate $sdf_7p_3$-shell.

From the panels of the energy levels of $\Nu{O}{18}{}$ and
$\Nu{F}{18}{}$, we can draw the same conclusion for the degenerate
$sd$-shell; the folded diagram  contribution makes the energy levels 
independent of the parameter $E$.

\subsubsection{Comparison of KK and EKK methods}\label{sec:sdf7p3_comp}

The main advantage of our EKK method is that
it allows for a fully consistent treatment of the non-degenerate model space. 
On the other hand, in the KK method, the naive perturbative calculation
of \Qbox by Eq.~\eqref{eq:Veff_Qbox_expansion} leads to divergence 
in non-degenerate $sdf_7p_3$-shell, as discussed in  Sec.~\ref{sec:EKK_MB}.

One possible {\it ad hoc} way to avoid this divergence is to artificially force the 
$sdf_7p_3$-shell to be degenerate in energy by shifting the single-particle
energies~\cite{Holt:2010yb}.
Although there is no physical justification for this artificial shift,
it removes the poles which arise in the various \Qbox diagrams.

Note, however, that the obtained $\Veff$ is defined
with artificially degenerate single-particle energies. 
Furthermore, in actual calculations, this method dose not necessarily 
lead to a convergent result even with all the folded diagrams in
Eq.~\eqref{eq:Veff_Qbox_expansion}.

For the purpose of comparing the EKK method with the {\it ad hoc}  treatment of the non-degeneracy 
in the
KK method, we apply the EKK method to both sets of  unperturbed
Hamiltonians;
one with the artificially degenerate $sdf_7p_3$-shell,
and the other with the non-degenerate $sdf_7p_3$-shell.
Note that the EKK method in the artificially degenerate $sdf_7p_3$-shell
simulates the {\it ad hoc} KK method explained above.
In this calculation, in order to obtain convergence, we set $E=-1.5$
MeV in both cases.
\begin{figure}[htbp]
 \begin{center}
  \includegraphics[width=9cm,angle=0,clip]{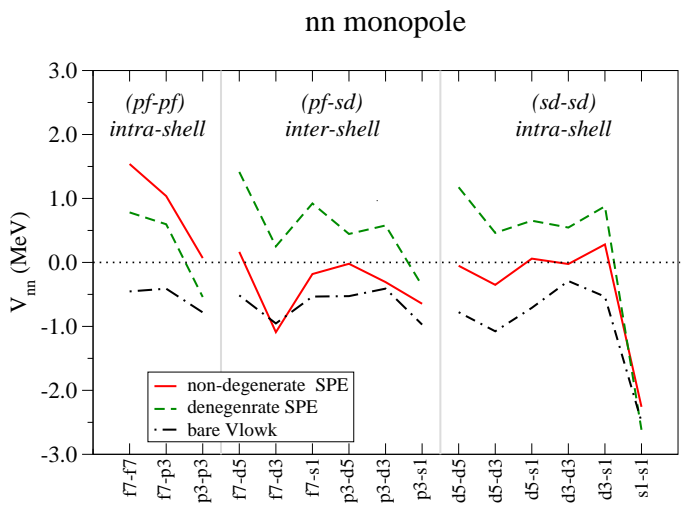}
 \end{center}
 \caption{(Color online)
 The monopole part of $\Veff$ in the $sdf_7p_3$ model space,
 obtained via the physical (non-degenerate) 
 single particle energies (full line)
 and via the {\it ad hoc} (degenerate) single particle energies (dashed line).
 Both calculations are performed in the neutron-neutron channel.
 As a reference, the monopole part of the first-order effective
 is shown with dash-dotted line.
}
 \label{fig:sdpf_diff_nn}
\end{figure}

Figure~\ref{fig:sdpf_diff_nn} shows the monopole part of the effective
interaction $\Veff$ in the $sdf_7p_3$-shell interaction for the
neutron-neutron channel.
The dashed lines show the results with the {\it ad hoc} modification of
the unperturbed Hamiltonian as explained above.
The full lines display the results in the non-degenerate model space.
To show the contribution of the \Qbox and folded diagrams, we display also
the monopole part of the first-order term of the \Qboxf. This term is energy independent.

It is interesting to note
that the inter-shell {\it (pf-sd)} interaction and 
the intra-shell {\it (sd-sd)} interaction are more attractive 
in the EKK method than in the {\it ad hoc} KK method.
This discrepancy clearly comes from the difference in the energy
denominator in the \Qboxf. 
Suppose we have two particles within $f_{7/2}$ or $p_{3/2}$ states.
The magnitude of the energy denominator shown in
Eq.~\eqref{eq:EKK_diag} is larger, making thereby the \Qbox smaller
 in the EKK method than in the {\it ad hoc} KK method.
For the inter-shell {\it (pf-sd)} interaction and 
the intra-shell {\it (sd-sd)} interaction, 
this difference makes the interaction
for the artificially degenerate single-particle states more repulsive.
On the other hand in the intra-shell {\it (pf-pf)} interaction,
the EKK results are more repulsive than the {\it ad hoc} KK results.
This can be understood as follows;
the effect of the folded diagram contribution is quite large in the EKK
method since  we use $P\tilde{H} P$ in the folding procedure
instead of $PVP$. This difference makes the contribution of the
folded diagrams larger.

The above comparison shows that there is a sizable difference between
the effective interaction in the EKK method and in the {\it ad hoc} KK
method,
and the difference affects mostly  the inter-shell
interaction. 

\subsection{\texorpdfstring{EKK method in $pf$-shell and
$pfg_9$-shell}{EKK method in pf-shell pfg9-shell}}
\label{sec:pf_pfg9}

We apply the EKK method to the nuclei
$\Nu{Ca}{42}{}$ and $\Nu{Sc}{42}{}$ as well. These nuclei can be described as two nucleons 
on top of a $\Nu{Ca}{40}{}$ core.
Here the degenerate model space is composed of  
 the single-particle states
of the $pf$-shell, while  our non-degenerate model space
is defined by the single-particle states of  the $pf$-shell and the single-particle state $g_{09/2}$. 
We have performed the calculation in the
same way as in Sec.~\ref{sec:sdf7p3}, but with an oscillator parameter
$\hbar\omega=11~\mathrm{MeV}$, which is appropriate for
$\Nu{Ca}{40}{}$ region.
 We have taken $\e{pf}=0$ and
therefore our first guess for the energy parameter is $E=0$.
\subsubsection{\texorpdfstring{Results for $pf$-shell}
{Results for pf-shell}}\label{sec:pf}
\begin{figure*}[htbp]
 \includegraphics[width=12cm,angle=0,clip]{fig8.eps}
 \caption{(Color online) $E$-dependence of the EKK results of neutron-neutron channel
 in the $pf$-shell (degenerate model space).
 (a) monopole part of the interaction, (b) level energies of
 $\Nu{Ca}{42}{}$ without folded diagrams and (c) with folded diagrams.
 Else, we use the same notation as in Fig.~\ref{fig:sd_result_nn}.
}
 \label{fig:pf_result_nn}
\end{figure*}

\begin{figure*}
 \includegraphics[width=12cm,angle=0,clip]{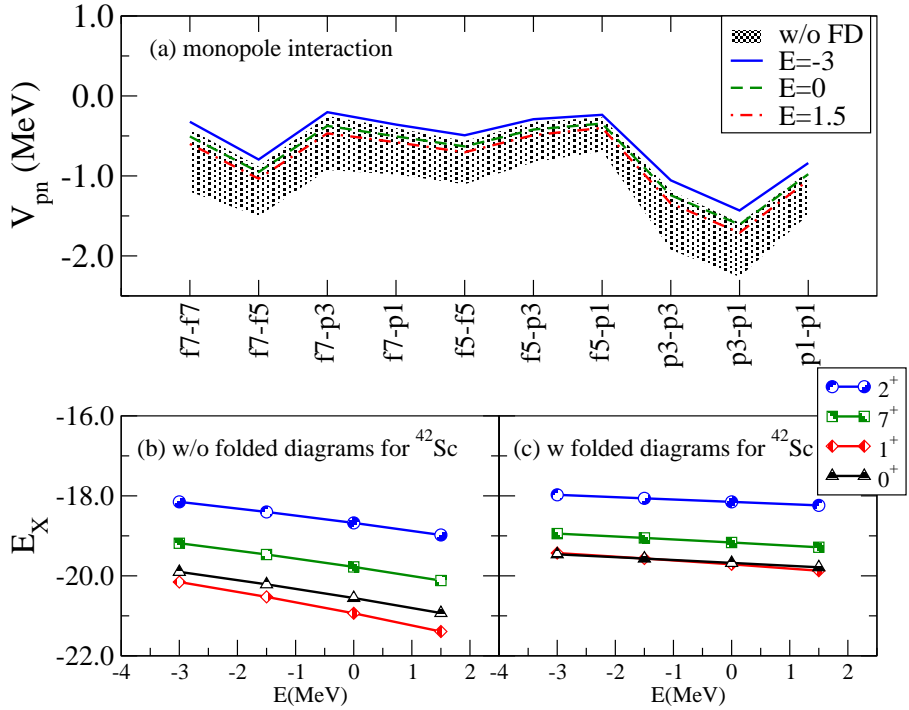}
 \caption{(Color online) $E$-dependence of the EKK results of proton-neutron channel
 in the $pf$-shell (degenerate model space).
 (a) monopole part of the interaction, (b) level energies of
 $\Nu{Sc}{42}{}$ without folded diagrams and (c) with folded diagrams.
 Else, we use the same notation as in Fig.~\ref{fig:sd_result_pn}.
}
 \label{fig:pf_result_pn}
\end{figure*}
Figure~\ref{fig:pf_result_nn},~\ref{fig:pf_result_pn} shows the
effective interaction $\Veff$ defined in the degenerate $pf$-shell. 
In Fig.~\ref{fig:pf_result_nn}, the panel 
(a) shows the neutron-neutron monopole interactions 
and the panels (b) and (c) display the energy levels of
$\Nu{Ca}{42}{}$. 
Figure~\ref{fig:pf_result_pn} 
shows the results for $\Nu{Sc}{42}{}$, the proton-neutron
channel, in the same manner. 
In the shell-model diagonalization,
 the single particle energies are taken 
from the GXPF1 interaction~\cite{PhysRevC.65.061301}; the particle
energies (in the isospin formalism) of $\e {f_{7/2}},~\e {p_{3/2}},~\e {p_{1/2}}$ and $\e
{f_{5/2}}$ are equal to $-8.6240$ MeV, $-5.6793$ MeV, $-4.1370$ MeV and
$-1.3829$ MeV, respectively.

We have carried out the calculation by varying $E$ in the range of
 $-3\leq E \leq 1.5~\mathrm{MeV}$.
Clearly, we can make the same observation as in Sec.~\ref{sec:sd};
the shaded bands in the monopole panels shrink 
when we include folded diagrams, making the monopole part of
$\Veff$ almost independent of $E$.
Moreover, from  the energies levels of $\Nu{Ca}{42}{}$
(Fig.~\ref{fig:pf_result_nn} (b) and (c)) and
$\Nu{Sc}{42}{}$(Fig.~\ref{fig:pf_result_pn} (b) and (c)),
we observe that the folded diagrams restore nicely the $E$-independence
of all the lowest-lying  levels, 
which is a natural consequence expected by the theory.
\subsubsection{\texorpdfstring{Results for $pfg_9$-shell}{Results for
   pfg9-shell}} 
\label{sec:pfg9}
\begin{figure*}[htbp]
 \includegraphics[width=12cm,angle=0,clip]{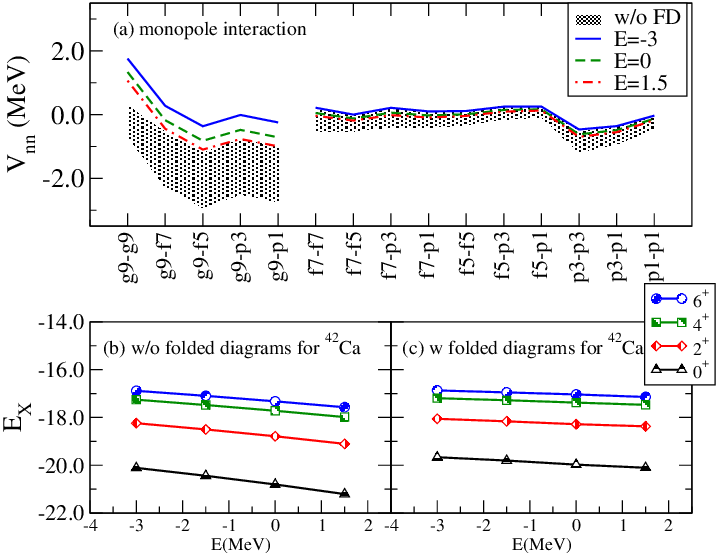}
 \caption{(Color online) $E$-dependence of the EKK results of neutron-neutron channel
 in the $pfg_9$-shell (non-degenerate model space).
 (a) monopole part of the interaction, (b) level energies of
 $\Nu{Ca}{42}{}$ without folded diagrams and (c) with folded diagrams.
 Else, we use the same notation as in Fig.~\ref{fig:sd_result_nn}.}
 \label{fig:pfg9_result_nn}
\end{figure*}
\begin{figure*}[htbp]
 \includegraphics[width=12cm,angle=0,clip]{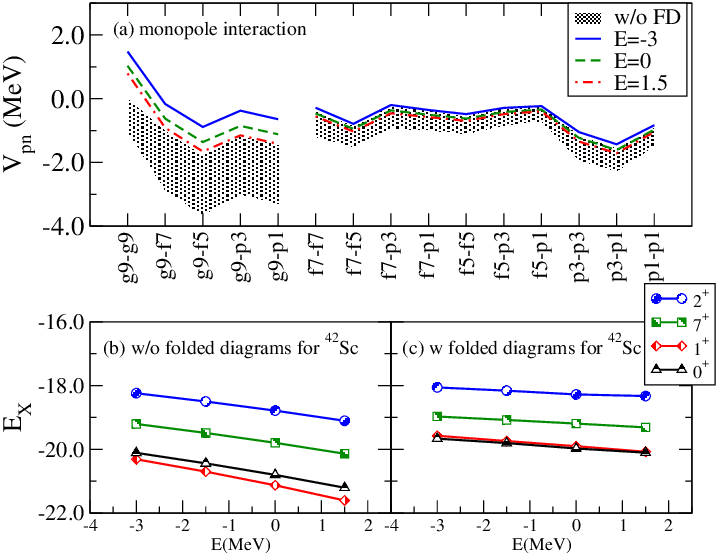}
 \caption{(Color online) $E$-dependence of the EKK results of proton-neutron channel
 in the $pfg_9$-shell (non-degenerate model space).
 (a) monopole part of the interaction, (b) level energies of
 $\Nu{Sc}{42}{}$ without folded diagrams and (c) with folded diagrams.
 Else, we use the same notation as in Fig.~\ref{fig:sd_result_pn}.}
 \label{fig:pfg9_result_pn}
\end{figure*}

As our final example, we present the results for the $pfg_9$-shell in
Fig.~\ref{fig:pfg9_result_nn} for the neutron-neutron channel and
Fig.~\ref{fig:pfg9_result_pn} for the proton-neutron channel.
The parameter $E$ is varied in the range of $-3\leq E \leq
1.5~\mathrm{MeV}$.
The particle energies (in the isospin formalism) of $\e {f_{7/2}},~\e {p_{3/2}},~\e {p_{1/2}},~\e 
{f_{5/2}}$ and $\e {g_{9/2}}$ are equal to $-8.6240$ MeV, $-5.6793$ MeV,
$-4.1370$ MeV, $-1.3829$ MeV and $2.1000$ MeV, respectively. 
We note again the role played by folded diagrams;
they remove the $E$-dependence of the $\hat{Q}(E)$, resulting
in an almost $E$-independent effective interaction $\Veff$ and 
energy levels of $\Nu{Ca}{42}{}$ (Fig.~\ref{fig:pfg9_result_nn} (b) and
(c)) and $\Nu{Sc}{42}{}$ (Fig.~\ref{fig:pfg9_result_pn} (b) and
(c)).


\subsection{Application of the EKK method to shell-model calculations}

With effective Hamiltonians derived by the EKK method, we can study
the role of such Hamiltonians in actual shell-model calculations.
Here we focus on the previously discussed systems, with at most two
valence nucleons outside a closed-shell core. Results from large-scale shell
model calculations will be presented elsewhere.


\begin{figure}[htbp]
 \includegraphics[width=8cm,angle=0,clip]{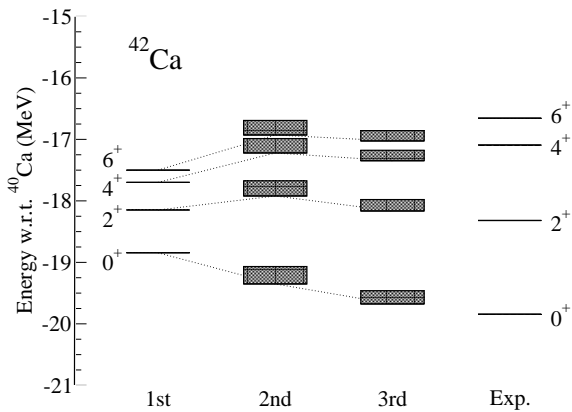}
 \caption{
 Energy levels of the ground state and low-lying states
 of $\Nu{Ca}{42}{}$ using  
 the \Qbox to first, second, and third order in the interaction.
  The right most levels represent the experimental values.
 Energies are measured from the closed-shell core, $\Nu{Ca}{40}{}$.
  The TBMEs (neutron-neutron channel) are obtained by the  EKK method, 
 with the parameter $E$ in the range $-3\leq E \leq 0$  MeV. 
 The shaded bands indicate the variation of calculated energy levels
 for $-3\leq E \leq 0$  MeV.
}
\label{fig:ca42}
\end{figure}

\begin{figure}[htbp]
 \includegraphics[width=8cm,angle=0,clip]{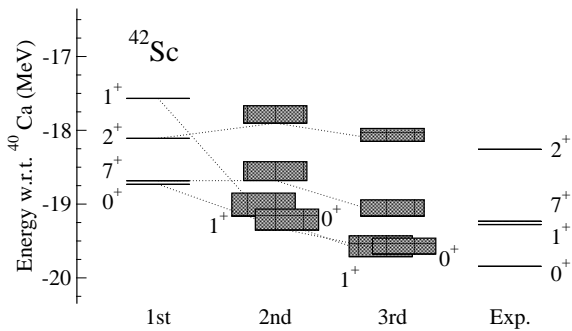}
 \caption{
 Energy levels of the ground state and low-lying states
 of $\Nu{Sc}{42}{}$ with respect to $\Nu{Ca}{40}{}$ core.
 The TBMEs (proton-neutron channel) are obtained via EKK method with
 parameter $-3\leq E \leq 0$  MeV. 
 Since we do not include Coulomb effects throughout the calculation,
 the absolute values are shifted by $-7.2$ MeV, 
 so that the $0^{+}$ state forms the $T=1$ triplet state 
 with the corresponding $0^{+}$ ground state of $\Nu{Ca}{42}{}$.
 Other notation is the same as for Fig.~\ref{fig:ca42}.
 }
\label{fig:sc42}
\end{figure}

For this purpose, 
Fig.~\ref{fig:ca42} shows the low-lying energy levels for
$\Nu{Ca}{42}{}$ obtained by the shell-model calculation with the same
setting as in Sec.~\ref{sec:pf}.
The TBMEs are derived by the EKK method, with the \Qbox 
calculated to first, second and third order in the interaction $V$.
The parameter $E$ is varied in the range of $-3\leq E \leq 0$ MeV in
the calculation of the second-order  and the third-order \Qboxf. Note that
the first-order \Qbox is $V$ itself, and is independent of $E$.
The leftmost levels show the result with the first-order \Qboxf.
The middle two levels represent the results with the second-order and third-order \Qboxf,
where the shaded bands show the range of the energy
levels corresponding to $-3\leq E \leq 0 ~\mathrm{MeV}$.
 The rightmost levels are the experimental data.

As discussed above, by construction, 
all the physical results are independent of $E$ if our \Qbox is
calculated without any approximation.
In other words,
 the weaker dependence on $E$ implies a better approximation  to the \Qbox
(see
Figs.~\ref{fig:sd_result_nn}-\ref{fig:sdf7p3_result_pn},
\ref{fig:pf_result_nn}-\ref{fig:pfg9_result_pn}).
This means that the width of the shaded band 
may be taken as a measure of the size of the error.

In Fig.~\ref{fig:ca42}, independently of the comparison to 
experiment, two types of convergence can be seen;
the first one is (i) the convergence of the energy levels  with
respect to the order of the perturbation,
and the second one is (ii) the convergence of the band widths
($E$-dependence of the energy levels) with
respect to the order of the perturbation.

Let us start with discussions on the convergence issue (i).
Figure~\ref{fig:ca42} shows that  
the magnitude of the third-order contribution 
is approximately 30\% as large as that
of the second-order contribution. 
A simple estimate from these observations tells us
that the perturbation series 
make a geometric series with common ratio $\sim 0.3$.
We can then expect that the fourth-order correction to the \Qbox are small.

Let us turn to the convergence case (ii).
Figure~\ref{fig:ca42} exhibits also that, compared with the second-order \Qboxf,  
the band widths are smaller for the third-order \Qboxf,
as we would expect.
Besides these theoretical features, 
the agreement with experiment  is acceptable.

Next we examine the proton-neutron channel using $\Nu{Sc}{42}{}$ as our testcase. The results are 
shown in Fig.~\ref{fig:sc42}.
The experimental values (right-most levels) are shifted by
$-7.2~\mathrm{MeV}$ so that the $0^{+}$ state forms the $T=1$ triplet 
state with corresponding $0^{+}$ ground state of $\Nu{Ca}{42}{}$,
since we presently do not take into account any isospin dependence of the 
nuclear force or the Coulomb interaction.

Here we observe similar patterns to those  seen in the neutron-neutron channel
of  Fig.~\ref{fig:ca42}.
However, we notice that as far as the convergence case (i) is concerned, the contribution of 
third-order terms is not
necessarily smaller than second-order contributions. 


On the other hand, for the convergence case (ii),
we observe smaller band widths for the third-order \Qbox 
results as compared with
 those obtained with the second-order \Qboxf.  
 This may indicate that the third-order \Qbox represents a better
approximation than the second-order \Qboxf.

The agreement with experimental data is rather nice for the $0^{+},
7^{+}, 3^{+}$ states, but we observe a slight overbinding for the
$1^{+}$ state.  From the convergence property of the $1^{+}$ state, it
is unlikely that fourth-order contributions to the \Qbox could play an
important role.  Several explanations for this discrepancy are possible.
One possibility for this overbinding is that the single particle
energies presently used are not fully adequate.  As mentioned above,
we have employed the single-particle energies determined from the
GXPF1~\cite{PhysRevC.65.061301} interaction. These energies are the
results of a fit together with the two-body interaction to
reproduce experimental data. The experimental data that enter the
fitting procedure consist of selected observables from $pf$-shell
nuclei.  Since the GXPF1 interaction differs from those derived here,
other sets of single-particle energies could have been more appropriate.
Another possibility is that our $Q$-space and order in perturbation theory may 
not be large enough. Thus, there may be
missing many-body correlations which could play an
important role.
Three-body forces \cite{PhysRevLett.105.032501,Holt:2010yb, Otsuka:2013ig, 
PhysRevLett.108.242501,PhysRevLett.109.032502} 
are examples of many-body contributions not studied here.
However, the aim of this work has been  to study the recently developed EKK formalism 
for deriving effective interactions for more than one major shell. The role of more
complicated many-body correlations are the scope of future works.


\section{Conclusion}
\label{sec:conclusion}

We have presented a novel many-body theory (the so-called EKK method) to calculate 
the effective interaction $\Veff$ for the shell-model calculation that
is applicable not only to degenerate model spaces, 
but also to non-degenerate model spaces.
The method is based on a re-interpretation of the unperturbed Hamiltonian and
the interacting part.
The final expressions for the effective interactions can be understood
as a Taylor series expansion of the Bloch-Horowitz Hamiltonian
around a newly introduced parameter $E$.
Since the change in $E$ should not affect the effective interactions
$\Veff$, the $E$-independence of the numerical results 
provides us with a sensible test of our framework  and the approximations
we make in the actual calculations.

In this work, we have presented numerical results 
for the effective two-body nucleon-nucleon
interactions $\Veff$, 
with applications to shell-model calculations of
selected nuclei, 
with and without the contribution from the folded diagrams.
The degenerate  model spaces are the $sd$-shell
for the nuclei $\Nu{O}{18}{}$ and $\Nu{F}{18}{}$,
and the $pf$-shell for the nuclei $\Nu{Ca}{42}{}$ and $\Nu{Sc}{42}{}$.
Our non-degenerate model spaces consist of 
the single-particle states from the $sdf_7p_3$-shell and the $pfg_9$-shell.
Based on our numerical results,
we have found that our method works well 
in practical situations.
For degenerate model spaces, our method
gives the same results as the conventional KK method.
In the non-degenerate model space, 
which is beyond the applicability of the KK method,
we have shown that our method works nicely. 

Second, we have shown that  
our $\Veff$ and therefore the energy levels 
are almost independent of the parameter $E$, as they should,
if we calculate the \Qbox up to third order.
This in turn suggests that the perturbative expansion of
the \Qbox through third order gives almost converged results.

Third,
the difference between the EKK method and the KK method 
with an {\it ad hoc}
modification of the unperturbed Hamiltonian is significant,
especially for inter-shell interactions.
This can have a large impact, for example, on   
the investigation of neutron-rich nuclei where 
the degrees of freedom defined by two or more major 
oscillator shells are important.

Finally, we stress that our method has
established a robust way to calculate microscopically the
effective interaction in non-degenerate model spaces. These spaces
involve typically more than one major oscillator shell.
We believe that this is an indispensable step to make
the nuclear shell model a reliable theory, in particular for exotic nuclei, 
based on a microscopic effective interaction
derived from the realistic nucleon-nucleon interactions.

\section{Acknowledgments}
We thank Gaute Hagen and Koshiroh Tsukiyama for fruitful discussions.
This work is supported in part by Grant-in-Aid for Scientific Research
(A) 20244022 and also by Grant-in-Aid for JSPS Fellows (No.~228635),
by the JSPS Core to Core program ``International Research Network for
Exotic Femto Systems'' (EFES), and the high-performance computing
project NOTUR in Norway. This work was supported by the Research
Council of Norway under contract ISP-Fysikk/216699.


\end{document}